# European ALMA operations: the interaction with and support to the users

Paola Andreani and Martin Zwaan

European Southern Observatory, Karl Schwarzschild strasse 2, 85748 Garching near Munich, Germany

## ABSTRACT

The Atacama Large Millimetre/submillimetre Array (ALMA) is one of the largest and most complicated observatories ever built. Constructing and operating an observatory at high altitude (5000m) in a cost effective and safe manner, with minimal effect on the environment creates interesting challenges. Since the array will have to adapt quickly to prevailing weather conditions, ALMA will be operated exclusively in service mode.

By the time of full science operations, the fundamental ALMA data product shall be calibrated, deconvolved data cubes and images, but raw data and data reduction software will be made available to users as well. User support is provided by the ALMA Regional Centres (ARCs) located in Europe, North America and Japan. These ARCs constitute the interface between the user community and the ALMA observatory in Chile.

For European users the European ARC is being set up as a cluster of nodes located throughout Europe, with the main centre at the ESO Headquarters in Garching. The main centre serves as the access portal and in synergy with the distributed network of ARC nodes, the main aim of the ARC is to optimize the ALMA science output and to fully exploit this unique and powerful facility.

The aim of this article is to introduce the process of proposing for observing time, subsequent execution of the observations, obtaining and processing of the data in the ALMA epoch. The complete end-to-end process of the ALMA data flow from the proposal submission to the data delivery is described.

Keywords: Instrumentation: submillimetre - millimetre facilities, Instrumentation: interferometry, Observatory: operations

## 1. THE ALMA FACILITY

When completed in 2012, the Atacama Large Millimetre Array (ALMA) will be the world's most powerful instrument for millimetre and sub-millimetre astronomy, providing enormous improvements in sensitivity, resolution and imaging fidelity in the mm and sub-mm bands. It is an aperture-synthesis array consisting of 54 twelve-metre and 12 seven-metre antennas, located at a 5 km high dry site in Chile, with a maximum baseline of 14.5 km. ALMA will provide enormous improvements in sensitivity, resolution and imaging fidelity in the mm and sub-mm bands, eventually covering all of the available atmospheric windows between 30 and 950 GHz in 10 bands with sub-milliJy sensitivity; six of these bands will be available initially.

The Array Operations Site (AOS) is located at the Chajnantor plateau at 5050m but operated from the Operations Support Facility (OSF) at 2950m near San Pedro de Atacama.

The array is a collaborative project between Europe, North America and East Asia, led by ESO, NRAO and NAOJ, respectively, and the competition for time from its global user community will be intense. European users in ESO member states will have access to observing time. Commissioning on the high site is expected to start in 2009, with an open call for proposals and Early Science in 2010 and full operation in 2012.

More about ALMA can be found in the ESO web page http://www.eso.org/sci/facilities/alma/about-alma/ and related links.

## 2. CONCEPTS OF ALMA OPERATIONS

From an astronomer's perspective, the ALMA Project Plan establishes several high-level concepts critical for science users: (1) *every astronomer, including novices to aperture synthesis techniques, should be able to use ALMA. (2) ALMA observations will be carried out in service mode and will be dynamically scheduled to optimally match the weather conditions and array configuration. (3) The calibration shall be reliable and self-consistent, so that data from the archive can be retrieved and reprocessed at any moment. (4) Data will be made public in a timely fashion.*

This implies that ALMA will be queue scheduled (i.e. Service Mode) all the time, the fundamental science deliverable will be calibrated and de-convolved images and ALMA will be accessible to a broad user community, from all branches of astronomy. An end-to-end science operations process will be put in place: easy-to-use software tools for proposal and observation preparation (supported by a group of knowledgeable observatory staff), an observatory-based plan for calibrating and monitoring system performance, a science data calibration and image production pipeline and an archive containing all associated science and technical information. Of course, as an observatory working at radio frequencies, ALMA will operate 24 hours a day – indeed, a provision in the ALMA design allows observations of the Sun.

To realize these ambitious goals three ALMA Regional Centres (ARCs) are established in Europe, North America and East Asia. These centres will be the access portals for the user community. Any interface between the user community and the ALMA observatory in Chile, from proposal submission to data delivery, will be happening through the ARCs. Indeed, for day-to-day operations, the three ARCs spread over three continents form an integral part of the overall ALMA operations. The ARC staff serves their regional communities, but also interact actively with Chilean observations. Science staff from the ARCs rotates through Chilean operations, providing the necessary close ties among the sites, and keeping the ARC staff familiar with the realities of observatory operations.

The relationship between the user, the ARC, and the Joint ALMA Observatory (JAO) in Chile, is schematically shown in Figure 1.

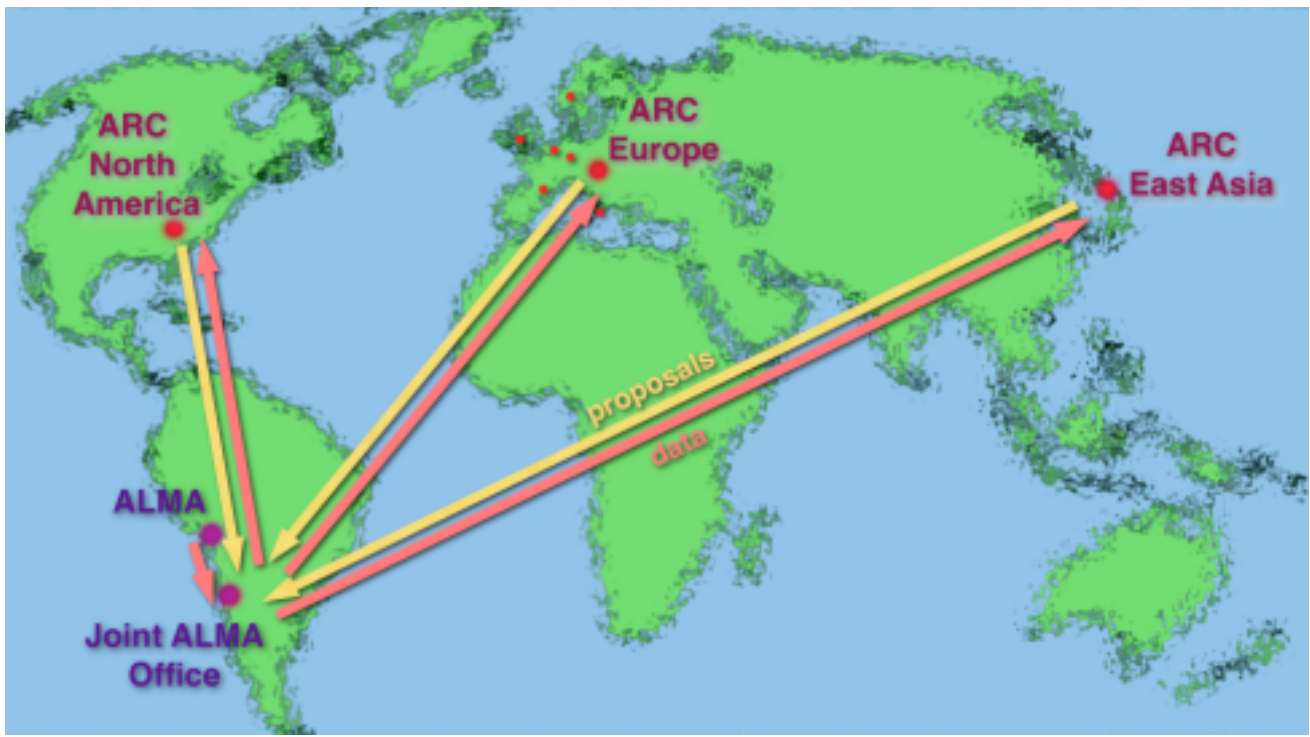

Figure 1: Proposals/Observing files are sent from the ARCs to the Joint ALMA Observatory (JAO) in Santiago (Chile). Data are sent from the JAO to ARCs by reverse route, with complete archives of all data at all four sites.

# 3. THE EUROPEAN ALMA REGIONAL CENTRE

In Europe, the ARC is currently set up as a network of nodes throughout Europe coordinated by a central node located at the ESO headquarters in Garching. This central node is part of ESO's Data Management Operations Division (DMO). The European (EU) ARC will be the point of contact for European ALMA users from the moment of proposal submission to the actual distribution of calibrated data and consequent analysis.

## 3.1 The central node

The core of the ARC activities at the central node will consist of assisting the user community with the technical preparation of observing proposals, ensuring that the observing programs are compliant with the requirements set by their scientific goals and make an efficient use of the facility, running a help-desk for the proposal submission and submission of observing programs, the delivery of data to principal investigators, the maintenance and refinement of the ALMA data archive, the feedback to the software development about data reduction pipeline and the off-line reduction software systems and all other operational critical software.

Readers may address themselves to the EU ARC web page, where more details about the EU ARC organization, tasks and news are continuously updated http://www.eso.org/sci/facilities/alma/arc.

## 3.2 The European ARC nodes

Fundamental to ALMA's success in Europe is the high level of user support services provided by the network of ARC nodes. These are required to fully realize the transformational nature of ALMA and to maximize the scientific return for the European community. The creation of non-ESO nodes recognises pre-existing expertise in single-dish and interferometric millimeter-wave astronomy elsewhere in Europe and a desire to build on that experience for the benefit of the entire European ALMA user community.

ARC nodes have been established in six different location and countries (in alphabetic order): Bologna (I), Bonn-Bochum-Cologne (D), IRAM (F, D, E), Leiden (NL), Manchester (UK), Onsala (SF, S, DK).

The principal role of the regional nodes is to provide face-to-face support for users from their national or regional areas, but they all have expertise in specific subjects which would benefit the wider European community.

Furthermore, fostering community development and guiding the future evolution of ALMA use, are among the nodes' primary tasks. The nodes will provide additional support, beyond what are called *the ARC core functions*, such as advanced user support for special projects and refinement in the data reduction process. In addition the sponsoring of workshops, schools, and events that stimulate the scientific activities around ALMA is very important for ALMA's visibility within the European programs of education and public outreach.

# 4. USER SUPPORT

## 4.1 Getting ALMA time

As usual, obtaining science data with ALMA begins with an observing proposal. ALMA observing proposals will be created and submitted using a software tool developed by ALMA but delivered and supported by the European ARC.
Once the Joint ALMA Office (JAO) issues calls for proposals, an astronomer wishing to apply for observing time will have to register on the ALMA web page. After registration, the user will make use of the ALMA Observing tool (AOT) to prepare a proposal. The AOT is a java application and is essentially a complete software package enabling one to construct a so-called *Observing Project* (see Figure 2 as an example). This Observing Project is the top item that any user will work on and consists of two parts: a *Phase I Observing Proposal* with emphasis on the scientific justification of the proposed observations and containing a minimal amount of technical information required to check the feasibility of the proposal, and a *Phase II Observing Program* submitted only if observing time has been granted. Once submitted, all proposals will be reviewed by a single Programme Review Committee. Accepted proposals will be ranked by scientific priority.

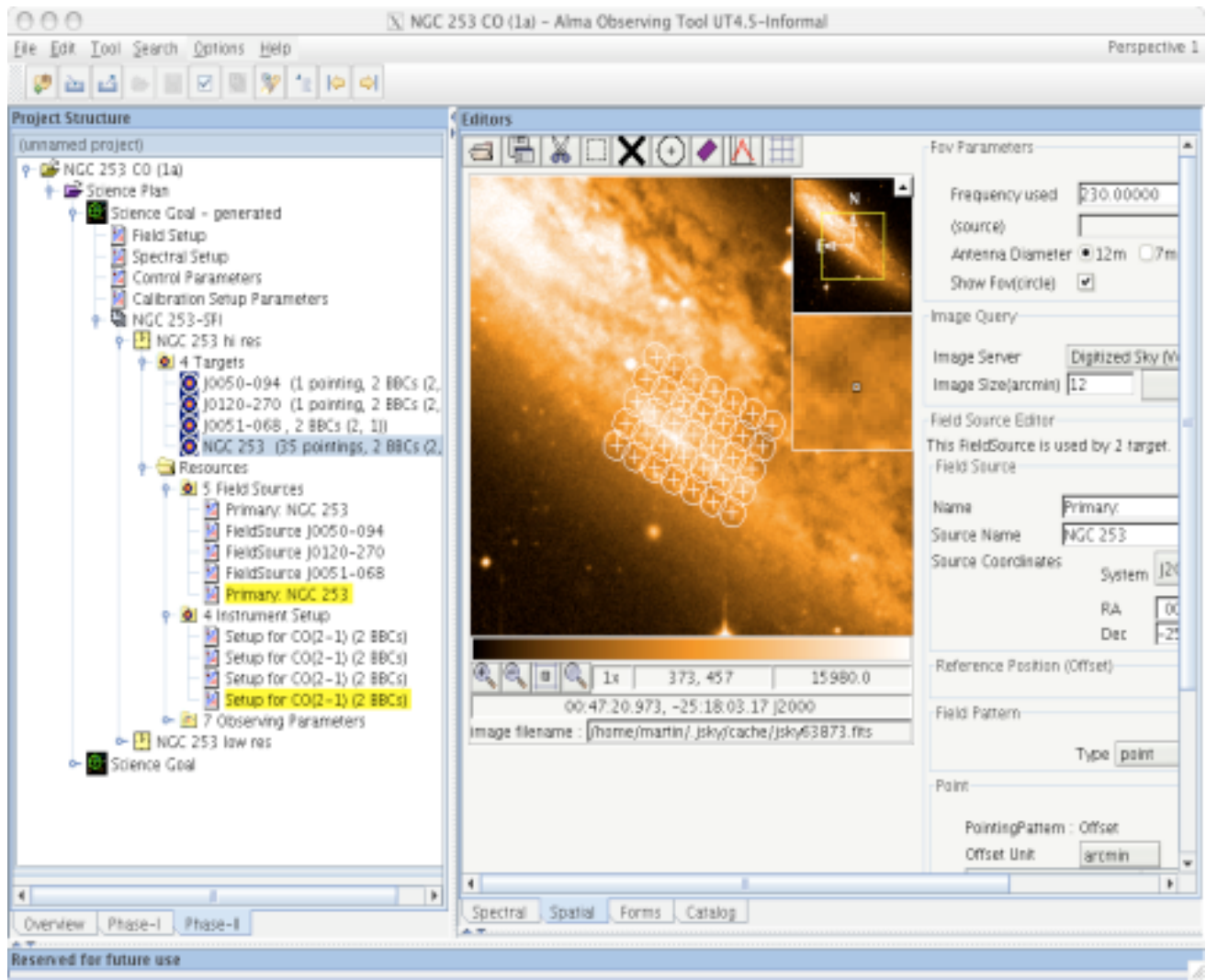

Figure 2: Display of the Visual Spatial Editor of the ALMA Observing Tool for an example mosaic observation of NGC 253. In the Editors pane the figure shows the image of NGC 253 as it was retrieved from the ESO Image server. The small circles represent the pointing positions for this target observation; the radius gives size of the primary beam at the observing frequency. On the left the project structure is visible.

The JAO, with assistance from the ARCs, coordinates the refereeing process. Users with accepted proposals in the merged schedule will then be invited to prepare a more detailed description of their desired observations in the form of Scheduling Blocks (SBs). The European ARC will assist the users during preparation of their Observing Project, providing documentation, proposal preparation and submission help.

In case the users require face-to-face help, they will be directed to their national or geographically closest ARC node,

unless it is a highly specialized issue, which can better be addressed at one of the other nodes. This face-to-face support is open to all users, although it is expected that normally users will utilise their nearest node (or the node funded by their own country or organization). Requests for specialized help will normally be directed to those nodes with expertise in the relevant subject areas.

With the use of the AOT the user needs to specify the technical details that control how the observations are to be carried out. The user creates a number of Scheduling Blocks (SBs) that contain all information necessary to execute a single observation. A Scheduling Block essentially consists of low-level observation commands to be submitted to the observing queue and will typically take 30 to 60 minutes to execute. It can be thought of as the smallest unit that can be scheduled independently. It is self-contained and usually provides scientifically meaningful data as well as a full description of how the science target and the calibration targets are to be observed. Sets of SBs can be combined with a description for the post-processing of the data, ultimately resulting in an image or a data cube.

The observing tool will allow astronomers, including those with little experience in aperture synthesis interferometry, to create full Observing Projects using standard observing modes. For more experienced users, who desire more control over the telescope configuration, more detailed specifications of each Scheduling Block can be given, such as the frequency setting of the local oscillator, the upper and lower side bands, the correlator parameters and the selection of base bands and sub band sets within each base band.

All material produced in this phase will be validated by ARC staff and released to the ALMA science operations team in Chile for scheduling and execution.

As with any queue-based system, SBs will wait in the queue until the ALMA arrays are in the desired configuration, atmospheric conditions are acceptable, the target is visible, and there is no other SB with higher scientific priority. When all these conditions are met, an SB will be executed. All antenna signals are correlated at the 5000-m AOS site. The *uv* data are transmitted via a fiber-optic cable to the OSF where these pass through a quick-look quality assurance pipeline. Based on the output of this pipeline and other technical telemetry, these data are judged to be valid or invalid by the ALMA astronomer on duty at the OSF. If user-related problems occur during SB execution, the ARC is notified and requested to work with the user to rectify the problems before the SB is queued again.

**4.2 Getting ALMA data**

ALMA observations will be dynamically scheduled, depending on weather conditions and the array configuration. Observations will be carried out 24 hours per day. Some projects may require only a single configuration, whereas others need observations using multiple configurations combined with ACA (Atacama Compact Array) and total power observations. Such a project may need several months to complete.

From the OSF, all data flow to the Santiago Central Office (SCO) where they are stored in the main ALMA archive and processed by the science grade data calibration and imaging pipeline. An ALMA astronomer on duty at the SCO inspects output images to assure *general* image quality, based on observatory-based goals, without trying to validate image quality requirements required for *specific* science programmes.

Before ALMA data reach the PIs, the data will pass through a multi-tier quality assurance program: a combination of on-site duty astronomer checks, a quick-look analysis, system performance checks and feedback from ARC staff. At the lowest level, science operations staff on duty at the OSF will monitor system performance (including the output from the on-site quick-look and telescope calibration pipelines) to detect periods when performance is out of the nominal range. To support this, the data from specific calibration SBs (e.g. flux standards) executed on a regular basis will be processed and checked against expected performance. At a higher level, the SCO science pipeline will produce images and associated QA information for review by ALMA science operations. Last but certainly not least, users will be encouraged to report the kind of subtle data problems only discovered by a detailed, science-based analysis. This feedback will be collected by the ARC staff and used as guidance to correcting problems or tuning system performance.

Images and associated quality assurance information are transmitted to the ALMA archive in Garching via the Internet. PIs will be notified immediately after their science data become available. The items made available to the PIs are the pipeline products (fully calibrated images or data cubes and calibrated *uv* data), raw *uv* plane source and calibration data, and offline data processing software including user support.

This process has several user and technical interfaces that must be managed diligently to achieve smooth operations and high user satisfaction. ESO has already a great deal of experience with managing such a transcontinental system and is working with its ALMA partners to apply "lessons learned". However, one technical challenge deserves highlighting: the coordination of five Petabyte class data archives at transcontinental distances. Some of this coordination occurs in near real-time over the Internet and some asynchronously as hard disks are shipped from Chile to the regional centers. Such a system has not been implemented on this scale before in astronomy.

It is essential to the success of ALMA that astronomers inexperienced in aperture synthesis imaging techniques are able to obtain science ready images and data cubes from their ALMA projects. The data reduction pipeline will therefore produce high quality science products for most standard observing modes. However, expert hands-on help will be required in many cases, especially when more complicated observing techniques are used. The first point of contact for data reduction help is the ARC main node in Garching, where users can address their questions through a helpdesk. Face-to-face help and any further specialized topics will be available from the nodes spread out over Europe (see section 3.2).

The offline pipeline data reduction software package responsible for generating science ready data products is CASA (Common Astronomy Software Applications). CASA has recently gone through major changes to optimize its use for ALMA data reduction. One of the most significant modifications is the creation of a completely re-designed python interface, which presents a very user-friendly way of interacting with the data reduction process. New functionality is continually added. Figure 3 shows an example of the CASA viewer.

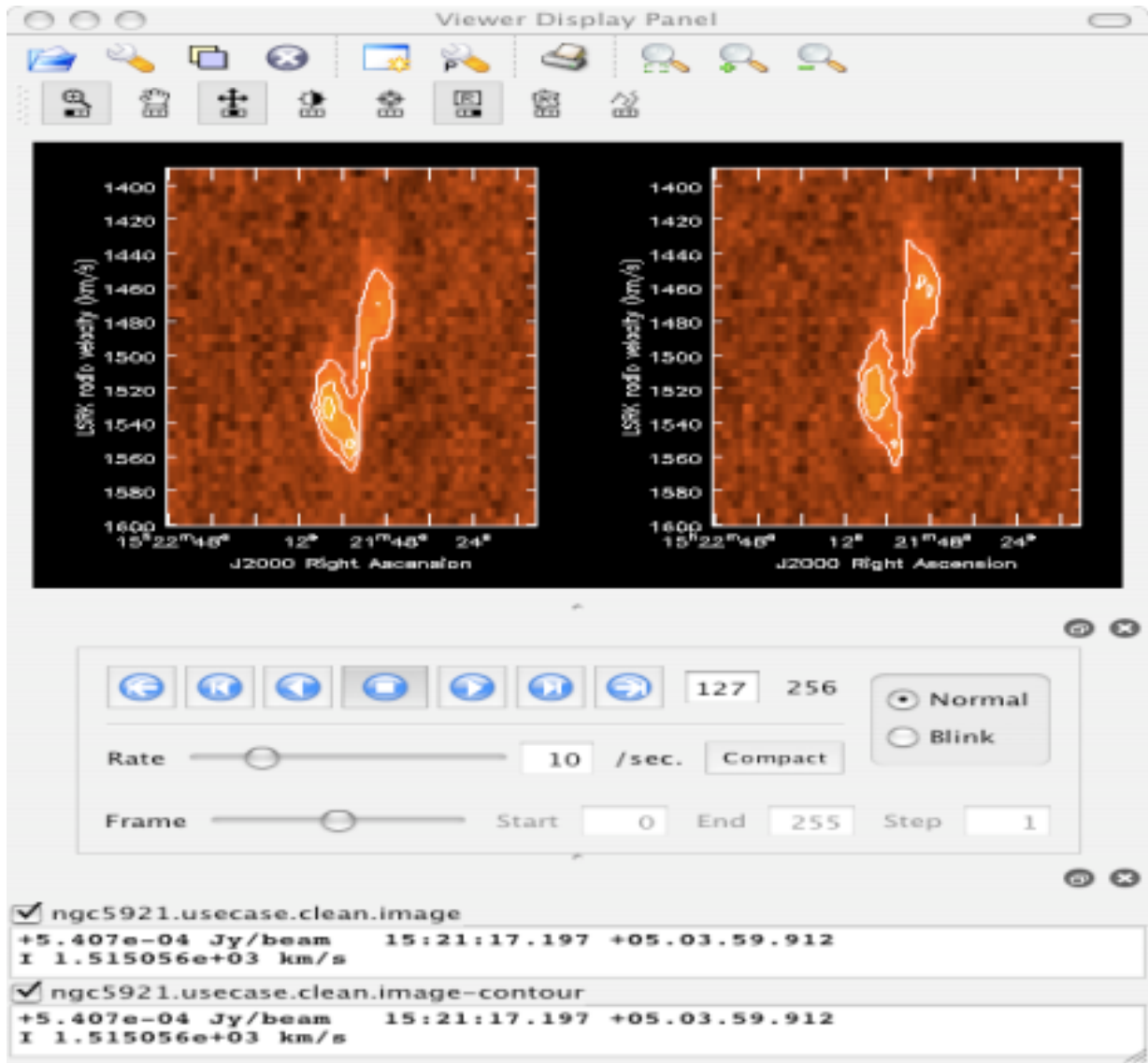

Figure 3: An example of the viewer display window in CASA. A position-velocity map is shown at two different positions. Contours are overlaid on a raster image. The viewer tool is very flexible as it can view (and edit) both visibility data and reduced data cubes.

Over the last few years, a series of user tests have been carried out to test the functionality of the data reduction software and to ensure that the development is adequate for ALMA needs. The tests have concentrated on many data reduction issues, and essentially covered the full end-to-end process from raw data sets to fully calibrated data cubes. CASA is currently in beta-release and will remain so until at least late 2008 or early 2009. Tests have now progressed from structured procedures using well-known data sets, to more realistic data reduction of science data. CASA is currently able to fill, edit, calibrate and image many continuum and spectral line data sets from a range of synthesis instruments. At present, CASA user support specialists are being trained in the US, East Asia and Europe. These specialists will train others in their region and provide support.

# 5. CONCLUSION

Although full ALMA operations will start in 2012, pre-operation activities have already started. The ARCs are organizing the support system, testing the software, writing cookbooks and manuals and preparing the commissioning and science verification phase, which will be starting in 2009. The first call for proposals for Early Science will be issued in early 2010 and the ARCs must be functioning at full speed before that date.

The international community can provide inputs into the ALMA project and operation through their representatives in the ALMA Science Advisory Committee (ASAC) and the European community through the European ALMA Science Advisory Committee (ESAC). Links to these committees can be found in http://www.eso.org/sci/facilities/alma/about-alma/org/asac.html and http://www.eso.org/sci/facilities/alma/about-alma/org/esac.html

Reference

P. Andreani & M. Zwaan, 2006, ESO Messenger 126, 43